\def\ct#1{\cite{#1}}

\def\somma#1#2{\sum_{\rm #1}^{\rm #2}}

\def\rfr#1{eq.(\ref{#1})}
\def\rfrs#1#2{eqs.(\ref{#1})-(\ref{#2})}

\def\Rfr#1{Eq.(\ref{#1})}

\def\eqi{\begin{equation}}
\def\eqf{\end{equation}}
\def\eqia{\begin{eqnarray}}
\def\eqfa{\end{eqnarray}}
\def\rp#1#2{{#1\over#2}}

\documentclass[11pt]{article}

\begin{document}

\noindent{\bf \LARGE{On the impossibility  of using the longitude
of the ascending node of GP-B for measuring the Lense-Thirring
effect}}
\\
\\
\\
 Lorenzo Iorio\\Dipartimento Interateneo di Fisica dell'
Universit${\rm \grave{a}}$ di Bari
\\Via Amendola 173, 70126\\Bari, Italy

\begin{abstract}
The possibility of analyzing the node $\Omega$ of the GP-B
satellite in order to measure also the Lense-Thirring effect on
its orbit is examined. This feature is induced by the general
relativistic gravitomagnetic component of the Earth gravitational
field. The GP-B mission has been launched in April 2004 and is
aimed mainly to the measurement of the gravitomagnetic precession
of four gyroscopes carried onboard at a claimed accuracy of $1\%$.
of better. The aliasing effect of the solid Earth and ocean
components of the solar $K_1$ tidal perturbations would make the
measurement of the Lense--Thirring effect on the orbit unfeasible.
Indeed, the science period of the GP-B mission amounts to almost
one year. During this time span the Lense-Thirring shift on the
GP-B node would be 164 milliarcseconds (mas), while the tidal
perturbations on its node would have a period of the order of
$10^3$ years and amplitudes of the order of $10^5$ mas.
\end{abstract}

\noindent Keywords: GP-B, Lense-Thirring effect, orbit, longitude
of the ascending node, orbital tidal perturbations

\section{Introduction}
The main scientific task of the Gravity Probe B (GP-B) mission
\ct{Everitt et al 2001}, launched in April 2004, is the
measurement of the general relativistic precession \ct{schi} of
four gyroscopes carried onboard induced by the Earth
gravitomagnetic field \ct{Ciufolini and Wheeler 1995}. The claimed
accuracy\footnote{See, e.g., the leaflet {\it Gravity Probe B
Launch Companion} downloadable from
http://einstein.stanford.edu/index.html} is of the order of 1$\%$
or better. The duration of the experiment is almost one year.

Measuring directly gravitomagnetism in a reliable and accurate way
would be a very important test of fundamental physics.

Up to now, the only attempts to measure it in the gravitational
field of the Earth have been performed by Ciufolini and coworkers
\cite{Ciufolini et al 1998, Ciufolini 2002}. They have tried to
measure the secular Lense--Thirring effect \ct{leti} on the whole
orbits of the existing laser-ranged geodetic LAGEOS and LAGEOS II
satellites by analyzing time series of the combined orbital
residuals of the nodes $\Omega$ of LAGEOS and LAGEOS II and the
perigee $\omega$ of LAGEOS II over time spans some years long. The
adopted  combination is not affected, by construction, by the
first two even zonal harmonics $J_2$ and $J_4$ of the static part
of the multipolar expansion of the Earth gravity potential whose
induced classical precessions severely alias the genuine
gravitomagnetic trends of $\Omega$ and $\omega$. The claimed
accuracy is of the order of 20-30$\%$ \ct{Ciufolini 2002}. For
different, more conservative but, perhaps, more realistic
estimates of the total error budget see \ct{Ries et al 2003, ior,
iormor}.
However, the multi-satellite approach could turn out to be more
fertile and fruitful in view of the improvements in the present
and near future Earth gravity field solutions from the CHAMP
\ct{Reigber et al 2004a} and, especially, GRACE \ct{Tap et al
2004, Reigber et al 2004b} dedicated missions. Indeed, they could
allow to greatly reduce the systematic error due to the even zonal
harmonics, so to discard the perigee of LAGEOS II and use the
nodes only. In \ct{iormor} a $J_2-$free combination which involves
the nodes of LAGEOS and LAGEOS II only has been explicitly
proposed: according to the recently released preliminary GRACE
models, it would allow a measurement of the Lense--Thirring effect
at a 15$\%$ level of accuracy (1 $\sigma$). Recently, a
combination involving the nodes of the geodetic LAGEOS, LAGEOS II
and  Ajisai satellites and of the radar altimeter Jason-1
satellite has been put forth \ct{iordoorn}: it would allow to
reduce the error due to the geopotential by cancelling out the
first three even zonal harmonics. The major problems could come
from the fact that the non--gravitational perturbations on
Jason--1 should be carefully dealt with. It must be pointed out
that the multi-year multi-satellite $J_{\ell}$-free approach
allows to perform, in principle, as many analyses as one wants
because of the extremely long lifetimes  of the satellites to be
used\footnote{The lifetime of the LAGEOS satellites is of the
order of $10^5$ years. LAGEOS and LAGEOS II are in orbit since
1976 and 1992, respectively. } and of the availability of data
records many years long. Moreover, since, in this case, one is
interested in the precise satellites' orbit reconstruction, such
analyses can benefit from the improvements both in the
observational techniques and in the modelling of the perturbing
forces acting on the satellites which will become available in the
future. So, the reliability and the accuracy of a measurement of
the Lense-Thirring effect based on such an approach is not fixed
once and for all but is increasing, at least up to certain level.
The launch of another LAGEOS--type satellite like the proposed
LARES \ct{Ciufolini 1986, Iorio et al 2002} or, more recently,
OPTIS \ct{optis} would further enforce the reliability and
precision of such a measurement, which could be pushed to the
1$\%$ level or, perhaps, even better.

%
It seems, then, legitimate to ask if it would be possible to
enlarge and enforce the significance of the GP-B gravitomagnetic
test by measuring also the Lense--Thirring secular precession of
its node. A previous treatment of this problem can be found in
\ct{peterson}. Apparently, this possibility sounds very appealing
because of the strictly polar orbital configuration retained
during the science phase and of the notable accuracy reached by
the most recent GRACE-based terrestrial gravity models. Indeed,
the competing secular nodal precessions induced by the even zonal
harmonics of the Earth gravitational field are proportional to
$\cos i$ \ct{ior}, where $i$ is the inclination of the satellite
orbital plane to the Earth equator assumed as $\{x,y\}$ reference
plane. Although GP-B is much more sensitive to the higher degree
multipoles of the Earth gravity field than, e.g., the LAGEOS
satellites\footnote{The semimajor axis $a$ of GP-B amounts to
7027.4 km, while $a_{\rm LAGEOS}=12270$ km. The classical nodal
precessions fall off as $R^{\ell}a^{-(\ell+3/2)}$.}, the fact that
$i_{\rm GP-B}=90.007$ deg together with the high accuracy of the
latest solutions of the Earth gravity field from GRACE should
constrain the systematic error due to the mismodelling in the even
zonal harmonics of the geopotential to an acceptable level. For
the combined impact of the departures from nominal polar orbital
configurations and of low altitudes see \ct{polares}. But, as we
will show, this is not all the story.
\begin{table}[ht!]
\caption{Relevant parameters for the calculation of the solid
Earth and ocean tidal perturbations for some selected tidal
constituents and orbital parameters of GP-B. The degree $\ell=2$
only tidal terms have been considered. The considered tidal
constituents are the zonal ($m=0$) lunar 18.6-year tide (055.565
in the Doodson notation) and the tesseral ($m=1$) solar $K_1$
tide. See \ct{iortid} and the references therein for the quoted
numerical values of the tidal parameters. For $G,GM, R,\rho$ and
$k^{'}_2$ the IERS values have been directly adopted \ct{iers}.
The proper angular momentum of Earth has been calculated as
$J=I\omega$, where the value of \ct{iers} for the Earth daily
angular velocity $\omega$ and of \ct{kinosh} for the adimensional
moment of inertia $I/MR^2$ have been adopted. The orbital
parameters of GP-B are those released in the leaflet $Gravity\
Probe\ B\ Launch\ Companion$.} \label{tidpar}
\begin{tabular}{llll}
\noalign{\hrule height 1.5pt} Symbol & Description & Value &
Units\\
\hline $G$ & Newtonian gravitational constant & $6.67259\times
10^{-8}$ & cm$^3$ g$^{-1}$ s$^{-2}$\\
$GM$ & Earth $GM$ & 3.986004418$\times 10^{20}$ & cm$^3$ s$^{-2}$\\
$R$ & Earth equatorial radius & $6378.13649\times 10^5$ & cm\\
$J$ & Earth proper angular momentum & $5.86\times 10^{40}$& g cm$^2$ s$^{-1}$\\
$\rho$ & Ocean water density & 1.025 & g cm$^{-3}$\\
$H_{2}^{1}(K_1)$ & Tidal height of the $K_1$ tide & 36.87012 & cm\\
$H_{2}^{0}(055.565)$ & Tidal height of the $18.6$-year tide & 2.792 & cm\\
$k_2(K_1)$ & Love number for the $K_1$ tide & 0.257 & -\\
$k_2(055.565)$ & Love number for the $18.6$-year tide & 0.315 & -\\
$\delta_{2, 1}(K_1)$ & Lag angle for the $K_1$ tide & -18.36 & deg\\
$\delta_{2, 0}(055.565)$ & Lag angle for the $18.6$-year tide & -56.29 & deg\\
$C_{2, 1}^{+}(K_1)$ & Ocean tidal height for the $K_1$ tide & 2.83
&
cm\\
$\varepsilon_{2,1}^{+}(K_1)$ & Ocean hydrodynamics phase shift &
320.6 & deg\\
$k_2^{'}$ & Load Love number & -0.3075 &-\\
$a$ & GP-B semimajor axis & $7027.4\times 10^{5}$ & cm\\
$e$ &GP-B eccentricity & 0.0014 & -\\
$i$ &GP-B inclination & 90.007 & deg\\
$n$ & GP-B Keplerian mean motion & $1.0717\times 10^{-3}$ &
s$^{-1}$\\
$P_{\Omega}$ & GP-B nodal period & $3\times 10^{10}$ (1136.746) &
s (yr)\\
$\Omega_0$ & GP-B longitude of the ascending node & 163.26 & deg\\
$\dot\Omega_{\rm LT}$ &GP-B nodal Lense-Thirring shift& 164 & mas
yr$^{-1}$ \\
\noalign{\hrule height 1.5pt}
\end{tabular}
\end{table}
%
\section{The impact of the static and time-varying part of the Earth gravitational field}
The secular Lense-Thirring precession of the longitude of the
ascending node $\Omega$ of a test mass freely falling in the
gravitational field of a central spinning mass with proper angular
momentum $J$ is \eqi \dot\Omega_{\rm LT}=\rp{2GJ}{c^2 a^3
(1-e^2)^{3/2}},\eqf where $G$ is the Newtonian gravitational
constant, $c$ is the speed of light in vacuum and $a$ and $e$ are
the semimajor axis and the eccentricity, respectively, of the
orbit of the test particle. The gravitomagnetic shift of GP-B
amounts to 164 milliarcseconds per year (mas yr$^{-1}$ in the
following). See Table \ref{tidpar} for the relevant parameters of
the Earth-GP-B system.
\subsection{The role of the even zonal harmonics}
In order to give a really pessimistic and conservative estimate of
the systematic error due to the mismodelling even zonal harmonics
of the geopotential, we will sum the absolute values of the
mismodelled classical precessions \ct{ior} induced by the whole
range of the mismodelled $J_{\ell}$ coefficients according to the
variance matrix of the GGM01C GRACE-based model \ct{Tap et al
2004} up to degree\footnote{For the LAGEOS satellites a
calculation up to $\ell=20$ is well adequate.} $\ell=48$. It turns
out that, at 1 $\sigma$ level
\begin{equation}\left\{\begin{array}{lll}
\delta\dot\Omega_{\rm geopot}&=& 0.8\ {\rm mas\ yr}^{-1},\\
{\delta\dot\Omega_{\rm geopot}}/{\dot\Omega_{\rm LT}} &=& 5\times
10^{-3}.
\end{array}\right.\label{geop}\end{equation}
If a root--sum--square calculation is performed by taking the
square root of the sum of the squares of the various mismodelled
classical precessions a relative error of $1\times 10^{-3}$ is
obtained.

Another source of potential bias when a single orbital element is
used is represented by the secular variations of the even zonal
harmonics $\dot J_{\ell}$. In \ct{Eanes and Bettadpur 1996} it has
been shown that an effective $\dot J_2^{\rm eff}\sim\dot
J_2+0.371\dot J_4+0.079\dot J_6+0.006\dot J_8-0.003\dot J_{10}...$
can be introduced. Its magnitude is of the order of $(-2.6\pm
0.3)\times 10^{-11}$ yr$^{-1}$. It turns out that its mismodelled
part would induce a secular drift of the GP-B node of $3.1\times
10^{-3}$ mas only in one year.
\subsection{The role of the solar $K_1$ tide}
As we will now show in detail, the major problems for a
measurement of the nodal Lense-Thirring shift with a polar orbital
geometry \ct{polares} comes from the classical time--varying nodal
perturbations induced by the solid Earth and ocean tides.

A given tidal constituent of frequency $f$ induces the following
solid Earth tidal perturbation on the node of a satellite
\ct{iortid} \eqi \Delta\Omega_f = \rp{g}{na^2\sqrt{1-e^2}\sin
i}\somma{\ell=0}{\infty}\somma{m=0}{\ell}\left(\rp{R}{a}\right)^{\ell+1}A_{\ell
m}\somma{p=0}{\ell}\somma{q=-\infty}{+\infty}\rp{dF_{\ell m
p}}{di}G_{\ell p q}\rp{1}{f_p}k_{\ell
m}^{(0)}H_{\ell}^{m}\sin\gamma_{\ell m p q f},\label{solida}\eqf
where $g$ is the acceleration of gravity at the Earth equator, $R$
is the Earth equatorial radius, $A_{\ell m}$ is given by \eqi
A_{\ell m}=\sqrt{\rp{2\ell+1}{4\pi}\rp{(\ell-m)!}{(\ell +
m)!}},\nonumber\eqf the quantities $F_{\ell m p}(i)$ and $G_{\ell
p q}(e)$ are the so called inclination and eccentricity functions
\ct{kaula}, $k_{\ell m}^{(0)}$ and $H_{\ell}^{m}$ are the Love
numbers and the solid Earth tidal heights, respectively, $f_p$ is
the frequency of the orbital perturbation given by \eqi f_p=(\ell
-2 p)\dot\omega+(\ell -2p
+q)\dot\mathcal{M}+m(\dot\Omega-\dot\theta)+\sigma, \eqf with
\eqi\sigma=j_1\dot\theta+(j_2-j_1)\dot s+j_3\dot h +j_4\dot p +
j_5\dot N^{'} +j_6\dot p_s ,\eqf
while the phase of the sinusoidal function is \eqi\gamma_{\ell m p
q f}=(\ell -2 p)\omega+(\ell -2p
+q)\mathcal{M}+m(\Omega-\theta)+\sigma t-\delta_{\ell m
f}.\nonumber\eqf
 The
angular variable $\mathcal{M}$ is the satellite's mean anomaly,
$s,h,p,N^{'}, p_s$  are the luni-solar mean longitudes, $\theta$
is the Greenwich sidereal time and the $j_k,\ k=1,..6$ are small
integers which can assume negative, positive or null values. They
are arranged in the so called Doodson number \eqi
j_1(j_2+5)(j_3+5).(j_4+5)(j_5+5)(j_6+5)\eqf by means of which each
tidal constituent $f$ is named. In it the integer $j_1$ classifies
the tides in long period or zonal ($j_1=0$), diurnal or tesseral
($j_1=1$) and semidiurnal or sectorial ($j_1=2$). Finally,
$\delta_{\ell m f}$ is the phase lag angle of the response of the
solid Earth with respect to the considered tidal constituent.

The ocean tidal perturbation on the node of an Earth satellite
induced by a tidal constituent of frequency $f$ is \ct{iortid}
\eqia \Delta\Omega_f&=&\rp{1}{na^2 \sqrt{1-e^2}\sin
i}\somma{\ell=2}{\infty}\somma{m=0}{\ell}\somma{+}{-}A_{\ell m f
}^{\pm}\times\nonumber\\
&\times& \somma{p=0}{\ell}\somma{q=-\infty}{+\infty}\rp{dF_{\ell m
p }}{di}G_{\ell p q}\rp{1}{f_p}\left[\begin{array}{c}
  \sin\gamma_{\ell m p q f}^{\pm} \\
  -\cos\gamma_{\ell m p q f}^{\pm}
\end{array}\right]^{\ell -m\ {\rm even }}_{\ell -m\ {\rm
odd}},\label{oceanica}\eqfa where $\pm$ indicates the prograde and
retrograde components, \eqi A_{\ell m f}^{\pm}=4\pi
GR\rho\left(\rp{1+k^{'}_{\ell}}{2\ell +1}\right)C_{\ell m
f}^{\pm},\eqf $\rho$ is the ocean water density, $k_{\ell}^{'}$ is
the Load Love number, $C_{\ell m f}^{\pm}$ are the ocean tidal
heights,
\eqi\gamma^{\pm}_{\ell m p q f}=(\ell -2 p)\omega+(\ell -2p
+q)\mathcal{M}+m(\Omega-\theta)\pm(\sigma t-\varepsilon_{\ell m
f}^{\pm}),\eqf
and $\varepsilon_{\ell m f}^{\pm}$ is the phase shift due to
hydrodynamics of the oceans.

We will deal with the long-period components averaged over one
orbital revolution ($\ell-2p+q=0$) because we have to compare
their effects with the secular Lense-Thirring effect. In
particular, we will focus our attention on the degree $\ell=2$
terms. For $\ell=2$, $p$ runs from 0 to 2, and so, in virtue of
the condition $\ell-2p+q=0$, $q$ assumes the values $-2,0,2$.
Since the only non vanishing eccentricity function of degree
$\ell=2$ is that for $p=1,q=0$, it follows that the condition
$\ell-2p=0$ is also fulfilled. The frequencies of the
perturbations are, in this case
\eqi f_p=\dot\Gamma_f+m\dot\Omega,\label{period}\eqf
with \eqi \dot\Gamma_f=(j_2 -m)\dot s+j_3\dot h+j_4 \dot p+j_5\dot
N^{'}+j_6\dot p_s.\label{freq}\eqf Since the science phase of GP-B
will be almost one year long, it is of the utmost importance to
see if there are some tidal perturbations, with relevant
amplitudes, which could resemble as superimposed linear trends on
the gravitomagnetic signal during such a time span. The most
powerful tidal constituents in affecting satellite orbits are the
zonal ($m=0$) lunar 18.6-year tide, with Doodson number (055.565),
and the tesseral $m=1$ solar $K_1$ tide with Doodson number
(165.555). For such tidal lines the inclination factor of the
perturbing
amplitudes are \eqia \rp{1}{\sin i}\rp{dF_{201}}{di}&=&\rp{3}{2}\cos i,\\
\rp{1}{\sin i}\rp{dF_{211}}{di}&=& -\rp{3}{2}\left(\rp{\cos^2 i
}{\sin i }-\sin i\right),\eqfa respectively. This implies that a
nearly polar orbital geometry will be affected mainly by the $K_1$
tide than by the 18.6-year tide. This is an important feature
because the period of the nodal perturbation induced by (055.565)
depends only on the luni-solar variables and amounts to 18.6
years. The situation is quite different for the $K_1$ tide.
Indeed, from \rfrs{period}{freq} it follows that its nodal
perturbation has the same period of the node of the satellite. For
GP-B it amounts to 1136.746 years. From Table \ref{tidpar} and by
assuming a time span of one year \rfr{solida} yields a nominal
shift of $\Delta\Omega(K_1)^{\rm solid}_{\ell=2}=2.01264\times
10^5$ mas. By assuming a 0.5$\%$ uncertainty in the Love number
\ct{iortid} one gets $\delta[\Delta\Omega(K_1)_{\ell=2}^{\rm solid
}]=1000.6$ mas. \Rfr{oceanica} yields a shift of
$\Delta\Omega(K_1)_{\ell=2}^{\rm ocean }=4.94785\times 10^5$ mas
after one year. By assuming an uncertainty of 3.8$\%$ in the ocean
tidal height \ct{Lemoine et al 1998}, one gets
$\delta[\Delta\Omega(K_1)_{\ell=2}^{\rm ocean }]=1.8801\times
10^4$ mas. The 18.6-year tide would not pose particular problems.
Indeed, the nominal amplitude of its nodal perturbation would
amount to 2.7 mas only; the error in the Love number at that
frequency is estimated to be \ct{iortid} 1.5$\%$, so that
$\delta[\Delta\Omega(055.565)]=0.04$ mas. These results clearly
show that the bias induced by the $K_1$ tide on the node of GP-B
would not allow to use it in order to measure the Lense-Thirring
effect. Such results lead to the same conclusions of
\ct{peterson}.
\subsection{A multi-satellite approach} At this point one could ask
if it would be possible to include the node of GP-B in some
multi-satellite $J_{\ell}$-free combination. The answer, also in
this case, is negative because of the polar geometry of its orbit.
As shown in \ct{polares}, the coefficients with which the
residuals of the orbital elements of a polar satellite would enter
some combinations tends to diverge for $i\sim 90$ deg. In the case
of GP-B one could consider, e.g., a $\delta\dot\Omega^{\rm
LAGEOS}+c_1\delta\dot\Omega^{\rm GP-B}$ combination. The
coefficient $c_1$ of GP-B would be equal to\footnote{The
systematic error due to the mismodelling in the even zonal
harmonics would amount to 50-100$\%$, according to GGM01C model.}
-398 so that the impact of the $K_1$ tide would be further
enhanced.
\section{Conclusions}
In this paper we have investigated the possibility of measuring
the gravitomagnetic Lense-Thirring effect on the orbit of a test
particle by analyzing also the nodal rate of the GP-B satellite in
the gravitational field of the Earth. It has been launched in
April 2004 and its main task is the measurement of the
gravitomagnetic precession of four gyroscopes carried onboard at a
claimed accuracy of 1$\%$. The main problems come from the fact
that the orbital perturbations induced by the solid Earth and
ocean components of the solar $K_1$ tide on the GP-B node would
resemble aliasing linear trends superimposed on the genuine
relativistic linear signal. Indeed, the observational time span
would be of almost one year while the period of such perturbations
is of the order of 10$^3$ years. Moreover, their amplitudes are
three orders of magnitude larger than the Lense-Thirring effect.
This drawback would be further enhanced by including the GP-B node
in some multi-satellite $J_{\ell}-$free combinations because the
coefficient with which it would enter them would be very large due
to the polar geometry of the satellite.


\end{document}